\documentclass[letterpaper,twocolumn,10pt]{article}
\pdfoutput=1
\usepackage{usenix,epsfig,endnotes}
\usepackage{times} % use it when more space is needed
\usepackage{graphics}
\usepackage{amsmath}
\usepackage{verbatim}%wonso:for source code printing
\usepackage{algorithm,algorithmic}
\usepackage{subfigure}
\usepackage{balance}
\usepackage[hyphens]{url} % hyphenate URL: url looks better
\usepackage{hyperref} % url link properly work
\usepackage{multirow}
\usepackage{hyphenat}

\usepackage[compact]{titlesec} % remove some space around titles

\begin{document}

\date{}

\title{\Large \bf 
Lamassu: Storage-Efficient Host-Side Encryption
}

\author{
{\rm Peter Shah and Won So}\\
NetApp Inc.
} % end author

\maketitle

\thispagestyle{empty}

\subsection*{Abstract}

Many storage customers are adopting encryption solutions to protect critical data. Most existing encryption solutions sit in, or near, the application that is the source of critical data, upstream of the primary storage system. 
Placing encryption near the source ensures that data remains encrypted throughout the storage stack, making it easier to use untrusted storage, such as public clouds. 

Unfortunately, such a strategy also prevents downstream storage systems from applying content-based features, such as deduplication, to the data.
In this paper, we present \emph{Lamassu}, an encryption solution that uses block-oriented, host-based, convergent encryption to secure data, while preserving storage-based data deduplication. 
Unlike past convergent encryption systems, which typically store encryption metadata in a dedicated store, our system transparently inserts its metadata into each file's data stream. 
This allows us to add Lamassu to an application stack without modifying either the client application or the storage controller.

In this paper, we lay out the architecture and security model used in our system, and present a new model for maintaining metadata consistency and data integrity in a convergent encryption environment. We also evaluate its storage efficiency and I/O performance by using a variety of microbenchmarks, showing that Lamassu provides excellent storage efficiency, while achieving I/O throughput on par with similar conventional encryption systems.

\section{Introduction}
\label{sec:intro}

Storage users are understandably sensitive to data security on shared storage systems. Adding encryption to an existing solution can help to address such concerns by preventing unauthorized parties from accessing the contents of critical data. One popular approach that seems to have quite a lot of traction is to encrypt data close to the application, or even inside an application itself. This strategy simplifies down-stream security by ensuring that data is in an encrypted state by default as it moves downstream through the stack. 
This strategy can take many forms, such as built-in application encryption, 
OS-based file system encryption or VM-level encryption~\cite{FileVault2:Online, VMWorkstation:Online, Zadok98cryptfs:a}.
We term any encryption that runs on the same physical hardware 
as the primary application \emph{data-source encryption}.
 
In general, existing data-source encryption solutions
interfere with content-driven data management features 
provided by storage systems --- in particular, deduplication. 
If a storage controller does not have access to the keys used to secure data, 
it cannot compare the contents of encrypted data to determine which sections, if any, are duplicates.

In this paper, we present an alternative encryption strategy that provides the benefits of upstream encryption
while preserving storage-based data deduplication on downstream storage. 
Based on these conflicting priorities, 
we name our system \emph{Lamassu}, 
after the Assyrian guardian deity that combines elements of several creatures. 
Our system builds upon existing work in convergent encryption~\cite{DouceurABST02,Keelveedhi:2013,Storer:2008:SDD} to enable deduplication of encrypted data, 
but extends it to provide its services in a manner transparent to both application and storage, without the need for dedicated metadata storage or additional files. 
This makes our system flexible and portable, allowing it to be self-contained, and 
greatly simplifying deployment in existing application environments. 
Our work also introduces a scheme for providing crash-tolerant data consistency 
in a convergent system, as well as a mechanism for verifying the integrity of data after a crash.

Lamassu preserves deduplication at the storage back end 
by using a message-locked, convergent encryption strategy~\cite{DouceurABST02} to secure data in a way that preserves block-equality relationships in the ciphertext. 
In such a scheme, data is encrypted using keys that are derived from the plaintext, 
thus the message is \emph{locked} under itself~\cite{BellareKR12}.
The actual cipher used to secure the data can be any standard encryption scheme, such as the Advanced Encryption Standard (AES).
By using this approach, any two users who have access to the same plaintext 
will deterministically arrive at the same ciphertext before storing it on the back end.
As a result, the storage system receiving that data will be able to identify and deduplicate redundant blocks, 
even though it is unable to decrypt them.

Convergent encryption provides strong security on data 
that has high min-entropy, where it is very difficult to guess at the contents of messages accurately. Unfortunately, real production data is often much less random than ideal data; there are often identifiable patterns that an outsider can exploit to guess at data contents with a much higher success rate than random guessing. 
As a result, this approach is vulnerable to the so-called 
\emph{chosen-plaintext attack}~\cite{Keelveedhi:2013, Zooko:2009}. 
In this attack, an adversary takes advantage of the nonrandom nature of 
production data 
by guessing the data rather than the encryption key.
If attackers guess correctly,
they can generate the matching key and verify their guesses
by generating ciphertext blocks that match the victim's blocks. 

Other work in this field has explored alternative defenses against the chosen-plaintext attack. For example, DupLESS~\cite{Keelveedhi:2013} 
provides a mechanism that uses a double-blind key generation scheme to allow an application host and a key server to cooperatively derive convergent keys. In the DupLESS scheme, the key server never sees the data to be encrypted, and the application host never has access to the secret keys stored on the key server. The disadvantage of that system is that each key generation operation requires multiple network round-trips between the application host and the key server, making it impractical for block-level operation.

We have chosen 
a relatively simple defense against the chosen-plaintext attack
by adding in a secret key to derive
the convergent key before using it for encryption ~\cite{Zooko:2009}. This mechanism is similar to the domain key used to derive keys in DupLESS, but in Lamassu, clients are permitted direct access to the secret key and generate their convergent keys locally. 
With this mechanism,
an attacker executing a chosen-plaintext attack 
needs to guess both the contents of the plaintext and the secret key in 
order to generate a matching convergent key, and to succeed. 

Lamassu instances that use different 
secret keys will produce different ciphertext from the same plaintext, 
and data across those instances will not be deduplicated.
On the other hand, 
if two (or more) clients share a single secret key, they can all read and write data to a shared storage system through Lamassu, and their shared data 
can be deduplicated by that system. 
In effect, a set of clients that share 
a single secret constitute both a security zone and a deduplication group. 
We collectively term a group of tenants that share a key an \emph{isolation zone}. 
The details of how this shared secret is implemented is discussed 
in further detail in \S\ref{sec:design}.

In order to retrieve Lamassu-encrypted data from storage, a user must have access to the encryption keys used to secure that data. 
Because message-locked encryption produces keys based on plaintext, 
it produces a large number of keys that must be fetched along with 
the ciphertext in order to retrieve data. 
This unbounded mass of keys presents a \emph{metadata
management problem} that is intrinsic to a message-locked encryption strategy.

Past solutions have managed this cryptographic metadata by storing keys 
alongside the encrypted file data~\cite{DouceurABST02,Keelveedhi:2013}, 
or by building a dedicated metadata store that stores the keys separately from the primary data~\cite{Storer:2008:SDD}. 
In both cases, the cryptographic metadata itself must be secured,
usually by means of 
either symmetric or asymmetric key encryption. 
Such solutions complicate the process of replicating or migrating encrypted data,
because the separated key information must be managed in parallel. 
For cases in which the cryptographic metadata is kept in a dedicated store, 
that functionality must also be replicated wherever the data is to be housed.
Providing full replication or migration capabilities may require either modification
to the underlying storage controller's facilities 
or the addition of external tools to provide those capabilities outside of the controller.

In contrast, Lamassu implicitly inserts the cryptographic metadata generated by encryption into the data stream for each file. 
In order to avoid polluting the file's data blocks with highly entropic key information, thus hindering deduplication, 
Lamassu places this data into reserved sections of the file. 
In effect, a predetermined fraction of the blocks stored at the storage controller
will be devoted to encrypted metadata, rather than file data. 
These encrypted metadata blocks 
are indistinguishable from random data,
and will not be deduplicated by the storage controller.

\paragraph{Contributions.}
To the best of our knowledge, 
Lamassu is the first system that achieves the following:
First, it provides strong data-source encryption that preserves storage-based block 
deduplication, without requiring modifications to the application or storage, and 
without requiring a dedicated metadata store to manage convergent keys;
second, by embedding cryptographic metadata inside encrypted files, Lamassu allows both data 
and metadata to be automatically managed by existing tools and storage features;
and third, our metadata structure provides a mechanism for maintaining consistency 
between a file's data and its cryptographic metadata.
In order to accomplish those goals, Lamassu uses these key techniques:
block-oriented convergent encryption, 
insertion of encryption metadata to the data stream,
efficient metadata layout and multiphase commit algorithm, 
and a built-in data integrity check mechanism.

The rest of this paper is organized as follows. 
In \S\ref{sec:design}, we will lay out the design of our system, including our threat model, detailed encryption strategy, and metadata layout. 
We will also describe our consistency and integrity model. 
\S\ref{sec:implementation} provides details on our prototype implementation, followed by \S\ref{sec:result} which shows our experimental results. 
Finally, we discuss related work and conclude in \S\ref{sec:relwork} and \ref{sec:conc}, respectively.

\section{Design}
\label{sec:design}

\subsection{Threat Model}
Our threat model is informed by the ones used by past secure deduplication
work~\cite{DouceurABST02,Keelveedhi:2013}, and by our own analysis of the level of security that could make sense in an enterprise environment. Our threat model takes the form of a series of explicit assumptions about 
the capabilities of a potential attacker, and of the hardware and software available in our 
expected deployment environment, as follows:
\begin{itemize}
\item 
We assume that the basic cryptographic primitives 
such as AES that we use,
represent an ``ideal'' encryption function and cannot be broken by attackers. We 
further assume that any potential attacker is aware of the encryption mechanism that we
have chosen and how it works.
\item 
We assume that data will be stored on an untrusted, shared storage system. We further assume that the shared system may behave as an \emph{honest-but-curious attacker}~\cite{Goldreich87}, attempting to read stored data, but not acting to maliciously destroy data. An example of such an environment might be a public cloud storage system, which can reasonably be expected to preserve stored data, but which must be prevented from viewing data contents.
\item 
We assume that the storage system will have full access to all data that it stores, but that it will not have any prior knowledge of the contents of encrypted data, or of the keys used to encrypt that data. 
\item 
We assume that the storage system stores data from multiple tenants, and that those tenants might not trust each other. We assume that these tenants might gain access to any data stored on the storage system, including access to data blocks that they are not authorized to access, such as through improperly applied access control.
\item 
We assume that the data-source systems that belong to a single trust domain may share secret information through some mechanism, such as a key server and KMIP
(Key Management Interoperability Protocol).
\end{itemize}

Convergent encryption, applied upstream of the storage system, effectively prevents that system
from reading the contents of the data. We assume that an attacker cannot compromise the key
manager shared by clients to gain access to their shared master keys. If that happens, the 
attacker can effectively read the data stored by clients sharing that trust domain. Note that it
would be feasible to adapt our system to use a double-blind key generation system that protects against that sort of attack, 
such as that described by Bellare et al.~\cite{Keelveedhi:2013} at the cost of reduced I/O performance. 
However, we have not pursued this option
due to the large performance overhead involved.

The work presented here focuses on protecting the contents of user data from an outside attacker, while preserving deduplication, but does not include protection for directory structure information. It should be possible to improve on this limitation by adding encryption for file and directory names in a future revision.

\subsection{Encryption}
\label{subsec:design_encryption}

The term \emph{convergent encryption} describes any encryption scheme that preserves the following property: Given a particular plaintext, it will always generate the same ciphertext. In every other 
respect, a convergent encryption scheme should share the same properties as standard encryption 
schemes. Lamassu exploits this property to enable deduplication of encrypted data by ensuring that identical plaintext blocks are stored as identical ciphertext blocks. This means that Lamassu exposes 
information about block equality to any potential outside observer, but does not expose any additional information about the data. Existing work on convergent encryption strategies discusses the cryptographic security of this approach~\cite{DouceurABST02,Keelveedhi:2013}. In general, larger block sizes reduce the granularity of information exposed to a potential attacker, and reduce the amount of information that can be gleaned from the pattern of blocks stored on disk.
 
\begin{figure}%[hb!]
\centering
\epsfig{file=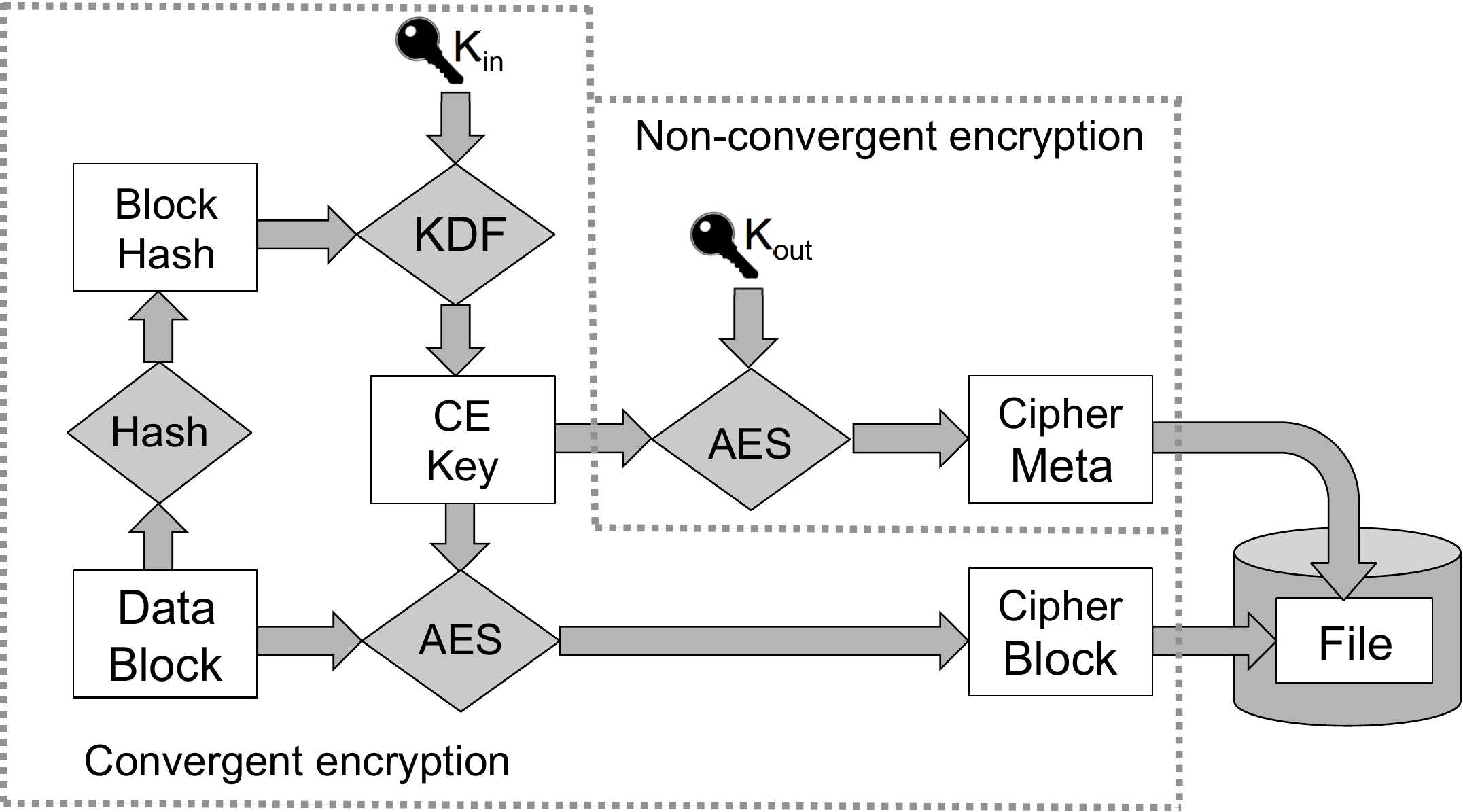, height=2.7in, angle=-90} %scale=.35}%width=\textwidth}
\caption{Lamassu's two-tier encryption model}
\label{fig:enc_model}
\end{figure}

Lamassu uses a two-tier encryption strategy, laid out in Figure~\ref{fig:enc_model}. The first tier is the convergent encryption applied to the application data written to each file. To protect against the chosen-plaintext attack, described previously, Lamassu uses a secret key in the process of deriving each convergent key. The second tier is standard (nonconvergent) encryption applied to the cryptographic metadata stored inside each file by using a second secret key. 
Data encrypted by separate Lamassu instances can be read or written by either instance, provided that those two instances share both of these secret keys.

The first of the two secret keys used by Lamassu is an inner key ($K_{in}$), used when encrypting file data blocks. When Lamassu writes a block to storage, it starts by taking a cryptographic hash\footnote{
Our current prototype is using SHA-256
in order to generate 32-byte hashes from fixed-size data blocks.} ($H$) 
of the data block in memory.
The convergent encryption key for that block ($CEKey$) is derived from the hash value 
and the inner key by the following equation:
\begin{eqnarray}
CEKey_i = F(H(Block_i), K_{in})
\label{eqn:blockkey}
\end{eqnarray}
where $F$ represents a \emph{key derivation function (KDF)}.
In our current implementation, this is accomplished by AES-encryption of the block hash using the inner key, but other key derivation functions could also work.
This modified $CEKey$ is used to encrypt the actual data block before it is sent to disk as shown in the following equation:
\begin{eqnarray}
CipherBlock_i = E_{AES}(Block_i, CEKey_i, IV_{fixed})
\label{eqn:cipherblock}
\end{eqnarray}
where $E_{AES}$ represents the AES encryption function.
For data block encryption, Lamassu uses AES-256 in CBC mode. As with previous convergent encryption systems, Lamassu uses a fixed initialization vector\footnote{Standard AES-CBC takes a randomly generated initialization vector (IV) as well as a key as inputs. Convergent encryption uses an invariant IV to preserve data equality in the ciphertext~\cite{DouceurABST02}.} ($IV_{fixed}$) for this process, so that future encryption of the same data will result in identical ciphertext~\cite{DouceurABST02}.
The block key is stored inside the file so that it can be easily retrieved 
when reading the file,
allowing Lamassu to decrypt the data block and recover its contents. 
The block key is stored in a reserved metadata section of the encrypted file. Further details will be described in \S\ref{subsec:metadata}.

Because the convergent keys are derived with key derivation function that uses a secret inner key ($K_{in}$), it is extremely unlikely that at encrypted block will match any data encrypted with the 
same technique, but using a different inner key. This property allows the inner key to be used to define a deduplication domain for data by restricting deduplication to just the data encrypted with 
the same inner key. In addition to preventing unauthorized parties from decrypting stored data, this also prevents an attacker from learning anything about secured data through the behavior of deduplication on multi-tenant, shared storage. This property allows tenants to define their own security isolation zone through the use of secret keys that are kept outside of the shared storage system.

The second shared secret used by Lamassu is an outer key ($K_{out}$) that is used to secure the metadata stored inside specially reserved sections of the file, including the per-block keys described previously. 
Lamassu encrypts the metadata blocks %, including per-block hash keys,
by using the AES in Galois/Counter mode (GCM), 
rather than in CBC mode as when encrypting data blocks.
Lamassu also seeds its metadata
block encryption with a randomly generated initialization vector ($IV_{rand}$)
like conventional encryption systems, as shown in the following equation:
\begin{eqnarray}
\begin{split}
CipherMeta_i
= E_{AES}&(Meta_i, K_{out}, IV_{rand})
\end{split}
\label{eqn:metablock}
\end{eqnarray}
where $Meta_i$ denotes a metadata block. 
A message authentication code (or tag) generated from AES-GCM
will be added to each metadata block and used for anintegrity check.
(The details will be discussed in \S\ref{subsec:consistency}.)

In order to read any of the data in the file, a Lamassu instance must have access to the outer key, thus defining a trust domain based on access to this key. Note that the outer key does not affect the boundaries of data deduplication, only data access. It would be possible to broadly share the inner key among many clients, while giving each one a separate outer key. 
The result would allow all of those clients to share a single deduplication isolation zone, while restricting them to reading and writing only their own private data.
However, note that cryptographic security among those clients would be exactly equal to that of basic convergent encryption. They would no longer have any protection against chosen-plaintext attacks executed by their peers.

The inner and outer keys dictate how Lamassu would approach periodic key rotation. Our experimental system does not include a mechanism to re-key Lamassu files, but it would be possible to approach the problem by rotating the secret keys stored in the key server. Key rotation would have to be initiated by a higher layer in the application with the ability to update the key server and to identify which files or directories need to be reencrypted with the new keys. An interesting side effect of Lamassu's encryption model is that it is possible to perform a less secure, but much faster partial re-keying of Lamassu data by changing the outer key, but not the inner key. In that case, only the metadata blocks in each file would need to be re-keyed, rather than entire files.

\subsection{Metadata Layout}
\label{subsec:metadata}

Lamassu's convergent encryption strategy operates on a per-block basis. This means that the base unit for any read or write is a full block. It is not possible for Lamassu to update a piece of a block without fully reencrypting the whole block with its new data. Furthermore, any change to a data block must be accompanied by a corresponding update to the hash key for that block in the metadata section of the file. 
We will discuss our strategy for maintaining consistency between data and metadata in \S\ref{subsec:consistency}.

Lamassu embeds its extra metadata into a file's data stream. This metadata is highly entropic by nature, and is extremely unlikely to result in 
any identifiable redundant sections for the storage controller to deduplicate. 
Because our metadata is produced in 32-byte sections rather than in full block-sized chunks,
writing the data into arbitrary sections of the file can pollute potentially deduplicable chunks, preventing 
them from matching other, similar blocks. It can also  interfere with block-alignment throughout the file, making it harder for a fixed-block deduplication to work.

Our solution is to place metadata in reserved sections of each file, segregating
cryptographic metadata from encrypted primary data completely. These sections are designed to
align with the underlying file system's block size so that they do not alter the block-alignment of any primary data. Our system is designed so that 
the chosen block size is easily variable. The chosen block size for our tests is 4096 bytes, with matching, aligned, 4096-byte reserved metadata sections inserted into the stream. This arrangement favors files that are at least a few megabytes in size, because this pre-allocation of space magnifies the space overhead 
of our solution in very small files. A smaller block size reduces the relative penalty for smaller files, but slightly increases the overall metadata space overhead for each file.

Because a file's size may be very large, and, more importantly, may change over time, Lamassu does not preallocate space for all of a file's metadata in advance. Instead, Lamassu distributes metadata blocks at regular intervals throughout the file, adding more as necessary. For simplicity, these blocks are placed in regular, predictable locations within the file, rather than in dynamically selected positions. Furthermore, each metadata block is placed in a position adjacent to the data blocks whose encryption keys it contains. We refer to a section of a file containing a single metadata block and all of the data blocks associated with it as a \emph{segment}.

\begin{figure}%[hb!]
\centering
\epsfig{file=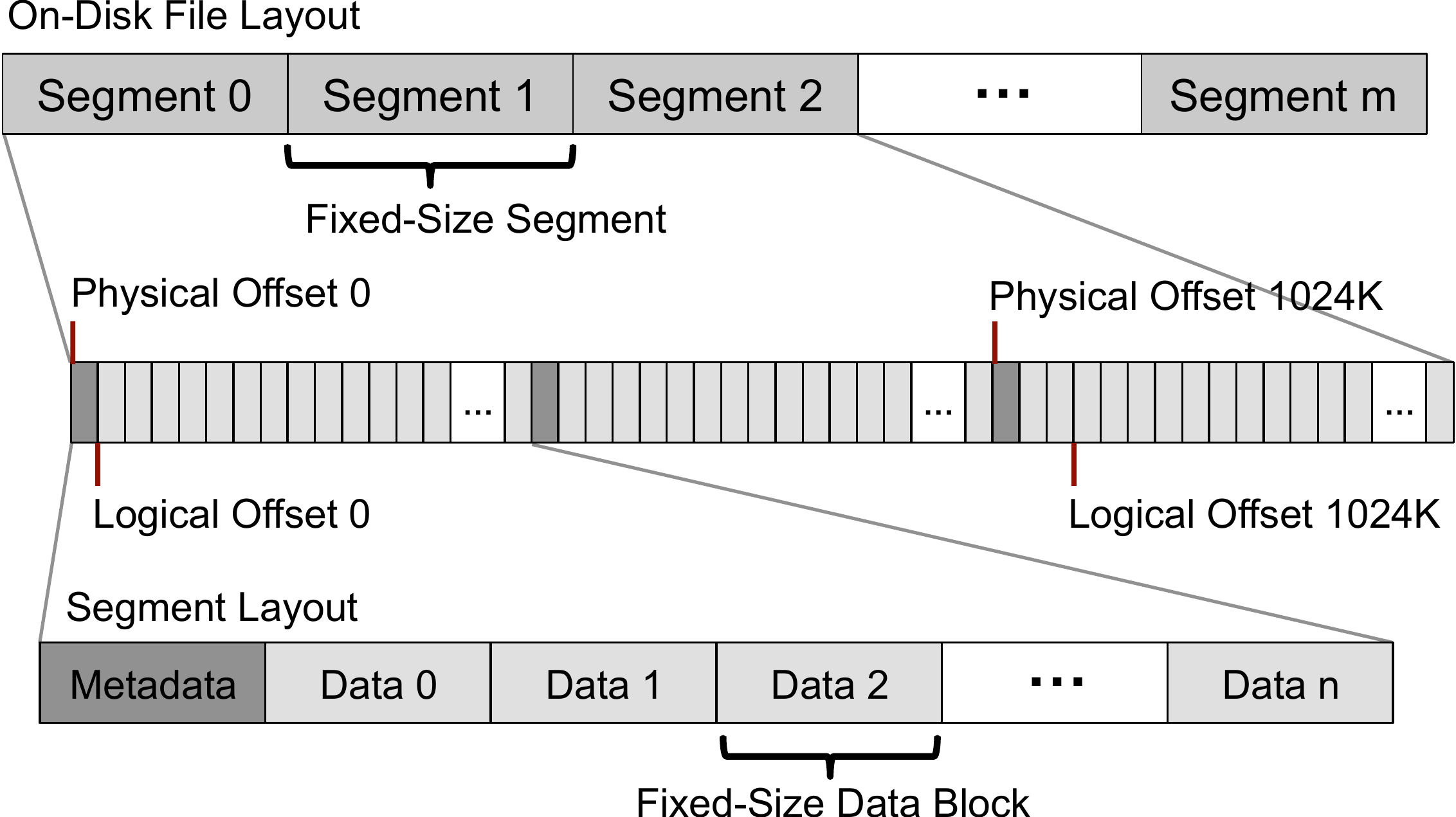, height=2.7in, angle=-90}%width=\textwidth}
\vspace{-.1in}
\caption{Internal layout of a Lamassu file}
\label{fig:file_structure}
\centering
\epsfig{file=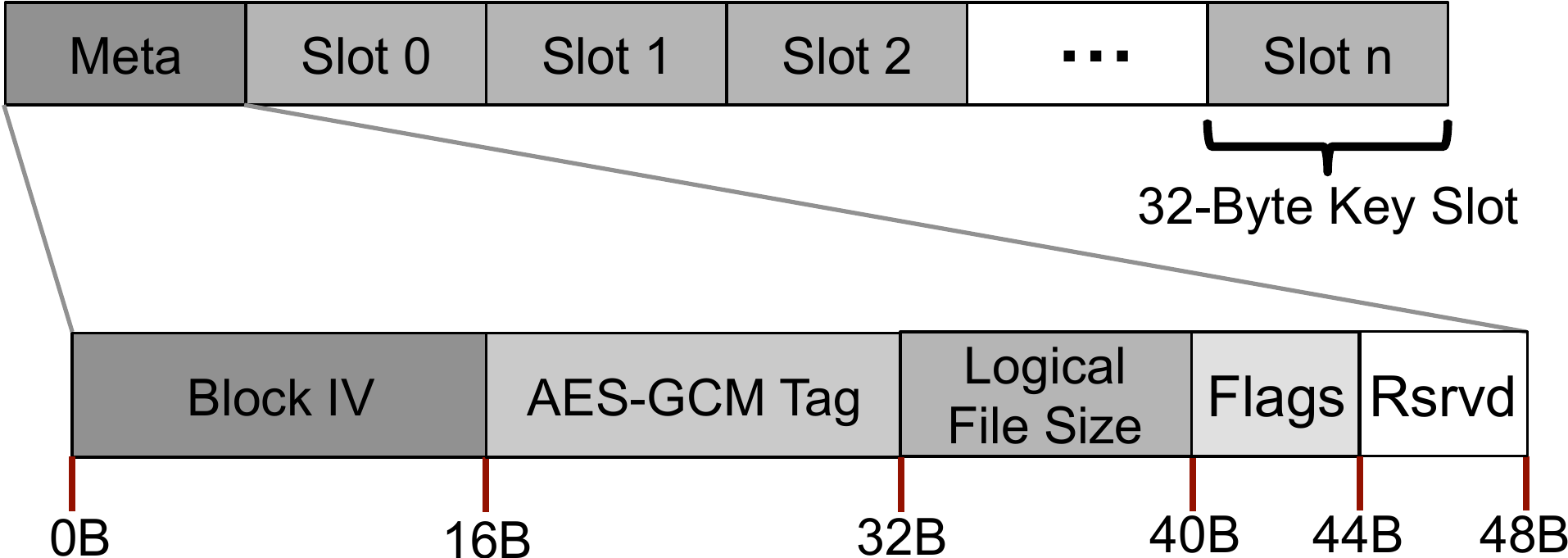, height=2.7in, angle=-90}%width=\textwidth}
\vspace{-.1in}
\caption{Internal layout of a Lamassu metadata block}
\label{fig:md_structure}
\end{figure}

Figure~\ref{fig:file_structure} shows the internal layout of a Lamassu file, based on a 4KB block size, with the file 
further broken up into smaller segments and blocks. The size of each segment is defined by the number of 32-byte encryption keys that can be stored inside a single metadata block. That number is affected by several factors, including the amount of space occupied by additional metadata information, and on tunable factors that are outlined in \S\ref{subsec:consistency}.

Inside each metadata block, the first 48 bytes of space are used for general file metadata, rather than for encryption keys. Figure~\ref{fig:md_structure} shows the contents of this metadata space,
which includes the random initialization vector (IV) used to encrypt the 
remainder of that block, 
a message authentication tag generated by AES-GCM,
and the logical size of the file's contents. 
The remaining space in each metadata block is taken up by a table of 32-byte encryption keys.

The size of each metadata block's key table determines the number of data blocks that can follow a single metadata block, and, by extension the size of a segment. Because each segment carries a mandatory one-block penalty for storing the metadata block, the most space-efficient arrangement is to maximize the size of each segment by filling as much space as possible with encryption keys. Lamassu trades away some of that space efficiency for better crash consistency, as will be discussed in 
\S\ref{subsec:consistency}.
When the size of plaintext data is $n$ bytes
and each metadata block can store up to $NumKeys_{MB}$ keys,
the number of data blocks ($N_{DB}$) and metadata blocks ($N_{MB}$), 
the size of the encrypted file ($n'$),
and the space overhead of Lamassu
can be formulated with the following set of equations:
\begin{eqnarray}
N_{DB} &=& \lceil n/BlockSize \rceil \\ 
N_{MB} &=& \lceil N_{DB}/NumKeys_{MB} \rceil \\
n' &=& (N_{DB} + N_{MB}) \cdot BlockSize \\
Overhead &=& n' - n %\frac{n'-n}{n}
\label{eqn:overhead}
\end{eqnarray}

The space overhead is minimized
when $n$ is exactly a multiple of the Lamassu block size
and the last metadata block has no empty key table slot,
as shown in the following equation:
\begin{eqnarray}
\begin{split}
Overhead_{min} = n / NumKeys_{MB}
\end{split}
\label{eqn:overheadmin}
\end{eqnarray}

As previously mentioned, the metadata stored at the beginning of each metadata block includes a logical file size. The reason for this is that Lamassu always encrypts data in full block-sized chunks, and, therefore, it both reads and writes data in full-block chunks. As a result, when writing to the end of a file whose size is not an integer multiple of the block
size, Lamassu will pad the final block with zeroes before writing it. 
In order to keep track of this padding and
report a correct file size to the application 
at a later time, 
Lamassu maintains the logical size of the file without this padding. 
(The logical size does not include the extra space taken up by key blocks stored inside the file.) This information is 
stored inside the Lamassu metadata blocks of the file.
Since it is highly inefficient to update every such block in a large file whenever the file changes, and it is always necessary to write to the file's last metadata block when changing its size, the updated size is written to the metadata block only for the final segment. The system always treats the file size stored in the final metadata block as the authoritative logical size for the file and ignores stale sizes that might be stored in any other blocks.

\subsection{Crash Consistency}
\label{subsec:consistency}
Lamassu's encryption model requires that the key for each data block be stored in the corresponding segment metadata block. Without a matching key, a data block cannot be decrypted, and becomes unreadable. This means that there is a critical failure mode wherein Lamassu crashes in between updating a data block in a segment and updating the metadata block for that segment, leaving the two in an unmatched, inconsistent state.

Lamassu addresses the threat of inconsistency due to incomplete writes 
by implementing a multiphase commit algorithm for writes. The sequence for each update is to first update the metadata block for 
the affected segment, and mark the segment as being in a midupdate state. When that has been completed, the modified data block is written out, 
and then, finally, the metadata block is re-marked to indicate that the 
update has completed.

To enable segment recovery after a failure, Lamassu stores extra key information in 
each metadata block during the update process. When Lamassu updates the segment metadata at the beginning of a data block write, it stores 
\emph{both} the new key \emph{and} the existing key for that block in the metadata block. Lamassu overprovisions the key table in each metadata block to provide space for a small number of transient, extra keys, stored during file writes. 
If the data block write succeeds, the subsequent update to the metadata 
block clears the update flag to indicate that the keys in the key 
table and the data in the data blocks are once again in sync.
Key table overprovisioning slightly reduces 
the number of keys stored in each segment, and consequently the amount 
of data in each segment, but we believe this to be a good trade-off for 
increased crash resiliency.

If Lamassu fails during the update process, it can recover based on the 
contents of the metadata block. If the system discovers a segment that is marked as midupdate, it can infer that a data block update was previously interrupted. If that is the case, it can detect which data block was in the middle of an update by reading the block number attached to the key, or keys, stored in the reserved space at the end of the key table. Once it has identified an affected block, it will 
be able to decrypt it using either the current key, stored in the 
key table, or the older key, stored in the reserved space, depending 
on whether or not the new version of the data block made it to 
disk before the crash.

Lamassu depends on the underlying storage system to provide consistency guarantees on whether or not a single block-level write reaches disk. 
Therefore, our method does not provide any mechanism for handling a partial-block write failure or disk write failures.

The penalty for the consistency model outlined above is an amplification in the number of disk I/Os that Lamassu has to perform whenever it updates a data block. To ameliorate this draw-back, Lamassu includes the ability to batch updates for multiple data blocks into a single update operation. To do this, Lamassu writes multiple keys to a metadata block as a single block update. 

Because each block included in the update must have two versions of its key written to the metadata block during the update, the number of blocks that can be combined into a single update is limited by the number of keys that can fit in the reserved space at the end of each metadata block. The precise amount of extra space reserved is adjustable at build time in our implementation. 
We use the parameter $R$ to represent the number of extra keys that can be stored in the reserved space for each metadata block.
Thus, with a single extra slot reserved (i.e., $R=1$), Lamassu will update a single data block at a time, requiring three I/Os for each block write: two for the metadata updates, and one for the data block itself. Increasing the number of reserved slots in each metadata block allows Lamassu to batch multiple data block writes into a single commit operation, amortizing the cost of the metadata updates across $R$ block updates.

Batching effectively reduces the system's I/O overhead.
However, reserving more key slots for old keys reduces the number of blocks that can be managed in a single segment, increasing the space overhead of Lamassu metadata. We will discuss the space and performance trade-offs introduced by varying $R$ 
in \S\ref{subsec:varying_R} with experimental results. Increasing $R$ also increases the amount of data that might be lost as a result of a midupdate crash.

\subsection{Data Integrity}
\label{subsec:integrity}
A useful property of Lamassu's encryption strategy is that it can automatically check whether the encryption key it uses to decrypt a data bock is the correct key for that block.
When Lamassu decrypts a data block by using a convergent hash key, it can immediately attempt to re-hash the decrypted block and recompute the hash key based on the resulting plaintext. If the plaintext is correct, the resulting hash key will match the one used in the decryption. If not, the resultant key is extremely unlikely to match the original hash key. Thus, a hash mismatch indicates a block-key mismatch. Lamassu 
takes advantage of this property to check the integrity of individual 
data blocks, checking the hash of decrypted data against the hash 
key stored inside the metadata blocks.

In the event of a crash and recovery, this hash-checking mechanism is 
what allows Lamassu to determine which of the two keys assigned to 
a data block matches the contents of that block. If Lamassu detects a
block-key mismatch that does not result from an interrupted write, it 
cannot correct the problem, but it can detect it and notify the client application.

Lamassu also includes integrity checking for metadata blocks, using AES-GCM authenticated encryption. 
AES-GCM attaches a message authentication code (MAC) to the encrypted metadata block. 
Decrypting the metadata block requires that the reader provide the MAC as well. In order to do that, the reader 
must already have access to the encryption key used to secure the block, and the secure hash of the block's original contents with which to verify its integrity.

Our design does not provide file integrity protection beyond the segment level. A malicious or defective storage system could, for example, roll the contents of a segment back from a current valid state to a previous valid state without having to read the contents of that segment. 
Our scheme would not detect such a change. To provide integrity checking at the level of a complete file, Lamassu would need to store data outside of the primary storage system, such as an on-premises store or, perhaps, in the key server. Lamassu's stackable design makes it possible to add an integrity layer on top of Lamassu, using a new, or existing, integrity checking system.

\section{Implementation}
\label{sec:implementation}

The Lamassu prototype system takes the form of a shim layer, 
sitting in the data path between the application and the back-end storage system.
Lamassu encrypts the data written by the application, inserts its metadata into the input data stream, and writes them all to the the backing store.
The precise amount of space overhead from metadata depends on 
the size of the files involved, and on the block size used.
Assuming the block size is 4096 bytes 
and that a single metadata block can store 125 keys per segment
(when $R = 1$),
the minimum space overhead ratio is 1/125 = 0.8\%.

We selected the Linux File System in User Space (FUSE)~\cite{FUSE:Online} 
as the infrastructure for our prototype. 
This arrangement allowed us to build our prototype 
as a self-contained user-mode program 
that can easily be ported into another application or infrastructure in the future.
Placing everything in a user space module also simplified our development 
and experimentation work.

\begin{figure}%[hb!]
\centering
\epsfig{file=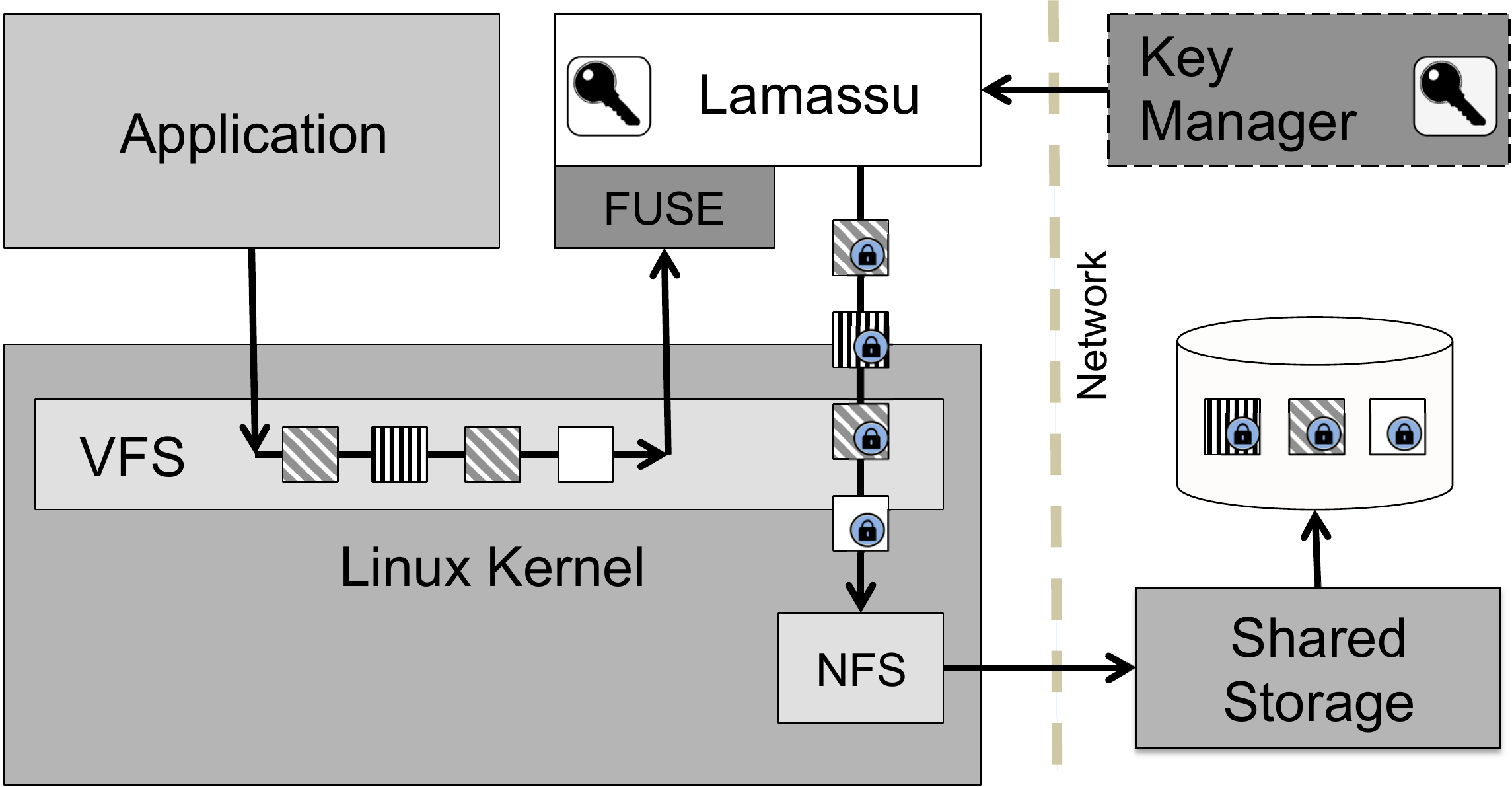, scale=.28, angle=-90}%width=\textwidth}
\caption{Lamassu prototype architecture}
\label{fig:prototype}
\end{figure}

Figure~\ref{fig:prototype} shows the flow of data through the system.
At start time, the Lamassu prototype selects a configurable directory, 
mounted on the native Linux file system, as its backing store. 
Lamassu will treat all files and directories in that mount point 
as Lamassu objects for it to manage. 
The underlying storage infrastructure for that directory can take any form, such as a local Linux file system, or an NFS mount point. 
Lamassu exports a file system interface to any Linux-resident application through FUSE and the Linux VFS layer. 
It accepts standard I/O requests and implicitly applies encryption, segmentation, and block chunking to each file 
before forwarding them to the backing storage system.
For most of our experiments, we used a NetApp\textsuperscript{\textregistered}
clustered Data ONTAP\textsuperscript{\textregistered} storage controller mounted over NFS as a deduplicating store.
Linux applications can access the encrypted file system 
through the Lamassu export by using standard file I/O interfaces. 

For key management, we used 
the Cryptsoft KMIP (Key Management Interoperability Protocol) SDK~\cite{KMIPSDK:Online}.
Two 256-bit AES encryption keys are retrieved at start time 
from a KMIP server:
One is used as an inner key ($K_{in}$), and the other is used as an outer key ($K_{out}$), 
as described in \S\ref{subsec:design_encryption}.
Every key created at the KMIP server contains an associated integer attribute
called an \emph{isolation zone}:
The clients in a single isolation zone
obtain the same set of encryption keys. This arrangement allows us to consistently match each Lamassu isolation zone to a KMIP isolation zone.

Our implementation exploits the architectural features provided by 
Intel processors to accelerate certain cryptographic operations. 
For the SHA-256 hash function, we make use of the Advanced Vector Extensions (AVX) instruction set by using an assembler library provided by Intel~\cite{Intel:SHA256}.
Where supported, our prototype also takes advantage of Intel's AES acceleration instruction set, AES-NI 
(Advanced Encryption Standard New Instructions), to maximize encryption performance. In such cases, the prototype uses AES-256 CBC encryption and decryption functions
provided by the Intel AES-NI Sample Library~\cite{Intel:AESNI}. On platforms where hardware acceleration is not available, our prototype defaults to the OpenSSL implementations for these functions.

\section{Experimental Results}
\label{sec:result}

\begin{figure}%[hb!]
\centering
\epsfig{file=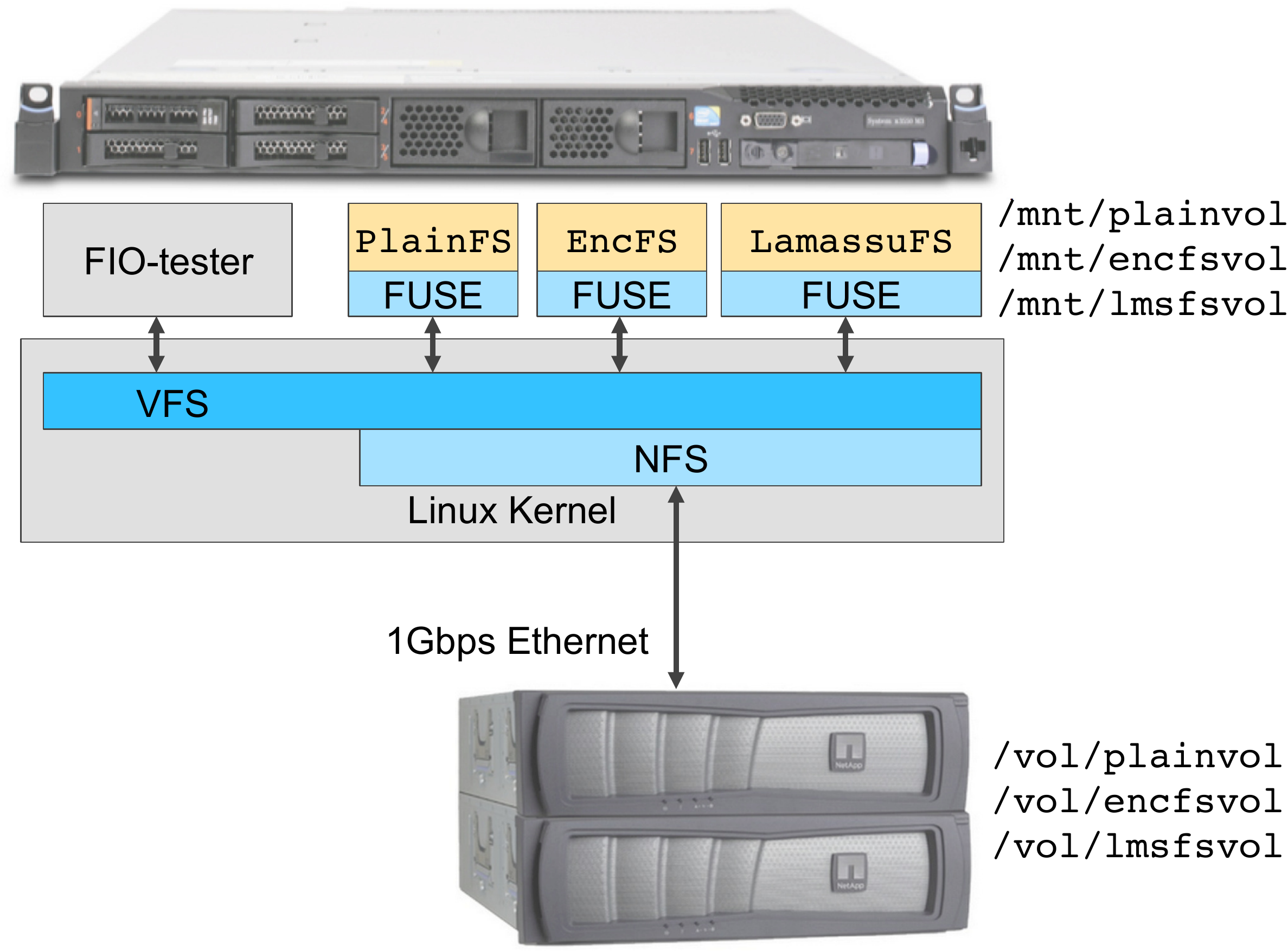, scale=.32, angle=-90}%width=\textwidth}
\caption{Experimental setup}
\label{fig:exp_config}
\end{figure}

\paragraph{Setup.}
For experiments, we set up an IBM server x3550
running 64-bit Linux (Fedora 20, Linux Kernel 3.3) as a host machine.
It has an Intel Xeon CPU E5-2630,
an 8-core processor supporting AES-NI,
which is critical for AES encryption/decryption performance.
The host machine is connected with a NetApp FAS3250 controller 
running clustered Data ONTAP 8
via a Gigabit Ethernet switch.
Figure~\ref{fig:exp_config} illustrates the experimental setup.

In addition to \emph{LamassuFS}, our FUSE-based Lamassu file system implementation,
we set up two additional file systems
that operate via FUSE.\footnote{FUSE version 2.9.3-2}
First, for a comparison with a conventional encrypted file system,
we chose \emph{EncFS}~\cite{EncFS:Online},
an open-source FUSE-based encrypted file system
that uses standard AES in CBC mode for encryption.\footnote{EncFS version 1.8-rc1}
Second, 
we also set up an unencrypted file system via FUSE,
which we refer to as \emph{PlainFS}.
This is mainly to provide
a fair comparison of performance 
against an unencrypted system that still inludes the FUSE overhead.
PlainFS is a simple pass-through front end for the 
relevant Linux system calls associated with FUSE operations.\footnote{
Most code from \emph{fuse-examplefs}~\cite{fuse-examplefs:Online}}

We created three separate volumes 
--- \texttt{plainvol}, \texttt{encfsvol}, and \texttt{lmsfsvol} ---
to be used as 
backing stores for PlainFS, EncFS, and LamassuFS, respectively.
Each volume is mounted on the host via \emph{NFSv3}
at a distinct mount point, and is used as a backing store
for the corresponding FUSE-based file system.
We used the same 4096-byte block size for both EncFS and LamassuFS.
This helped to keep our comparisons fair, and ensured that I/O operations 
for both would be aligned with the native block sizes of our
storage controller.
For most experiments, 
the number of reserved key slots in the metadata block ($R$)
was fixed to 8. 
With this setup, 
a single segment is composed of one metadata block followed 118 data blocks,
and the minimum amount of space overhead is 0.85\%.

\subsection{Storage Efficiency}
\label{subsec:res_se}

We first evaluated Lamassu storage efficiency to make sure that we could achieve the deduplication goals we had set. 
To do this, we wrote a simple tool to generate
4GB synthetic data files
with various redundancy profiles 
(as the percentage of redundant 4KB blocks in a file, denoted $\alpha$) 
ranging from 10\% to 50\%.
Each data file was copied over NFS to different volumes
in the storage system through PlainFS, EncFS, and LamassuFS.
When the copy completed, we manually triggered deduplication on the 
storage system.
We measured the difference in disk space usage before and after deduplication
using \texttt{df}, run on the controller itself.

\begin{figure}%[hb!]
\centering
\epsfig{file=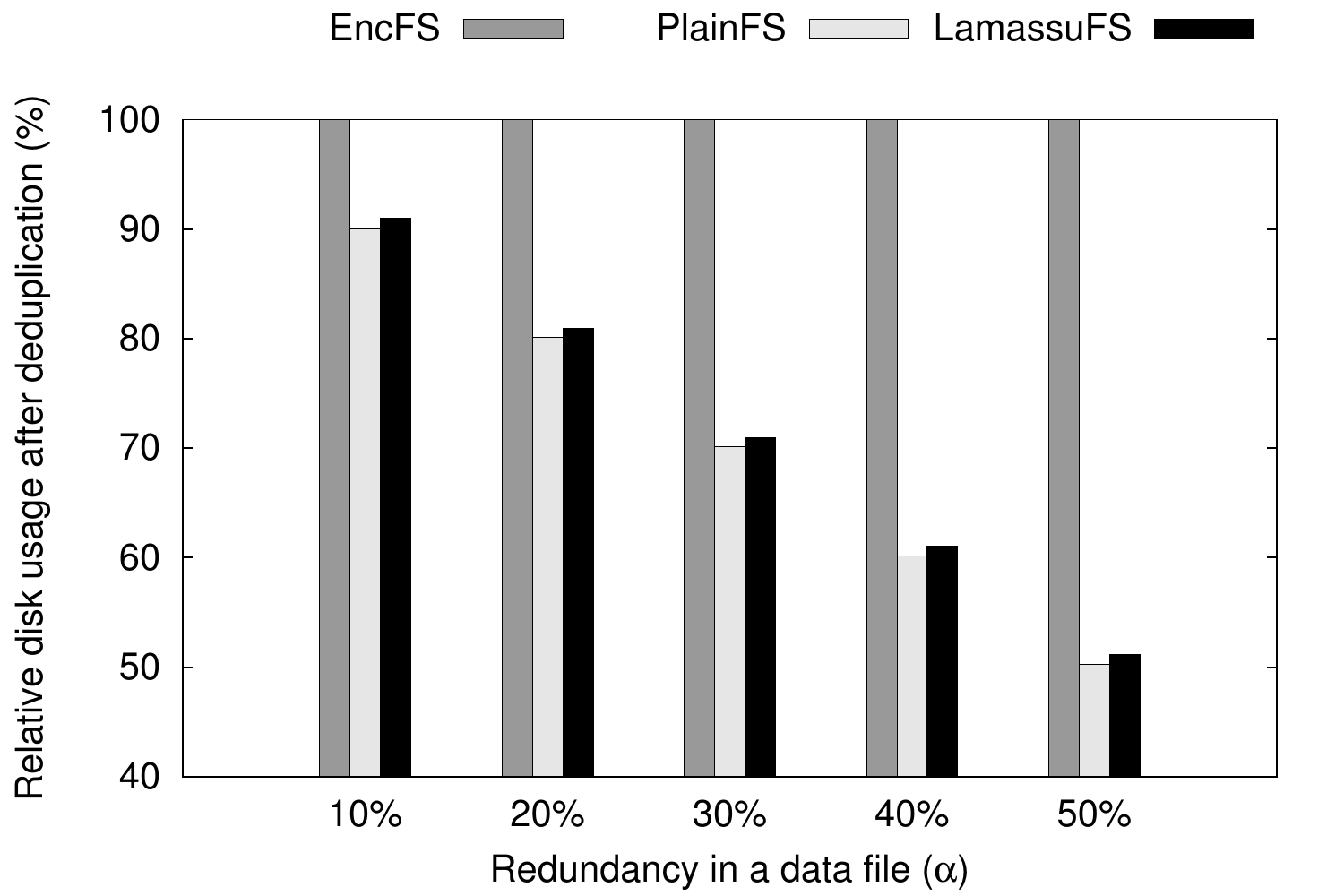, scale=.55, angle=0}%width=\textwidth}
\caption{Storage efficiency with synthetic files}
\label{fig:stor-eff}
\end{figure}

The relative percentage disk usage after deduplication 
is plotted in Figure~\ref{fig:stor-eff}.
EncFS shows 100\% for all cases 
because no deduplication occurred by using standard AES encryption.
For PlainFS where data files are stored as unencrypted blocks, 
the relative disk usage is exactly 
$(1 - \alpha)$. 
These results match our expectations, with the space savings from 
deduplication on unencrypted data mapping 1-to-1 with the known level of data redundancy on the test data. 
LamassuFS achieved nearly the same storage efficiency as PlainFS,
but with a small amount of space overhead due to the embedded cryptographic metadata.
This overhead is constant, relative to the nondeduplicated size of a file, 
but the relative overhead on deduplicated storage increases 
as the data file redundancy ($\alpha$) increases:
1.01\%, 1.06\%, 1.21\%, 1.43\%, and 1.81\% respectively,
i.e., inversely proportional to ($1 - \alpha$).

\begin{table}%[b!]
\caption{Storage efficiency with VM images}
\centering
\scriptsize 
\begin{tabular*}{\columnwidth}{@{\extracolsep{\fill}} | l | r | r | r | c |}
\hline
\multirow{2}{*}{VM image} & \multirow{2}{*}{Size} & \multicolumn{2}{c|}{\% Deduplicated}   & Space \\
\cline{3-4}
& & PlainFS & LamassuFS & overhead \\
\hline
FreeDOS.vdi & 379M & 9.35\% & 9.18\% & 1.07\% \\
FreeBSD-7.1-i386.vdi & 1.8G & 15.40\% & 15.11\% & 1.35\% \\
xubuntu\_1204.vdi & 2.3G & 22.07\% & 21.95\% & 1.01\% \\
Fedora-17-x86.vdi & 2.6G & 36.73\% & 36.46\% & 1.83\% \\
opensolaris-x86.vdi & 3.5G & 8.08\% & 7.87\% & 1.14\% \\
\hline
\end{tabular*}
\label{tab:stor-eff-vm}
\end{table}

The same set of experiments was performed by using real virtual machine images\footnote{Obtained from \url{http://virtualboxes.org/images/}} with various sizes 
as shown in Table~\ref{tab:stor-eff-vm}.
Note that EncFS results have omitted because they were all zero.
The storage efficiency results with real files are completely consistent with the results from synthetic data shown in Figure~\ref{fig:stor-eff}:
LamassuFS achieves almost the same amount of deduplication as PlainFS, 
with a small amount of space overhead of less than 2\%.

\subsection{Performance}
\label{subsec:res_perf}

Because AES encryption and decryption are a compute-intensive jobs,
using encryption in a file system incurs a performance overhead.
In order to examine this, 
we evaluated the I/O performance of PlainFS, EncFS, and LamassuFS.
For a fair comparison,
we carefully chose the EncFS configuration parameters:
4096 bytes for a block size,
AES-256 in CBC mode for an encryption algorithm,
and no file name encryption.
We also turned off all EncFS features that 
insert metadata between blocks. This change caused EncFS to write data in a block-aligned pattern, similar to Lamassu's. We did this because we have
observed that EncFS performs quite poorly when allowed to write in an
unaligned pattern.
EncFS also uses AES-NI through the OpenSSL library on platforms that support it.

In order to examine the performance of Lamassu 
under a larger variety of circumstances,
we used \emph{FIO-tester}~\cite{FIO:Online},
which generates various types of synthetic workloads
to all 3 file systems.
We applied 5 different workloads to a single 256MB file
with 4KB-block synchronous I/O:
sequential reads (seq-read), sequential writes (seq-write),
random reads (rand-read), random writes (rand-write), 
and mixed random reads/writes with the read/write ratio of 7:3 (rand-rw).
The I/O throughput (bandwidth) was
measured through 10 runs;
the Linux kernel page cache was flushed before each run
so that no data block was cached at the host memory.
For LamassuFS, one more variation was added: 
\emph{LamassuFS(meta-only)}, where the read path only
checks the integrity of metadata blocks
without checking the integrity of data blocks.
This would give a good indication of
how much performance is penalized when
providing a full data integrity check to the system.

\begin{figure}%[ht]
\centering
\epsfig{file=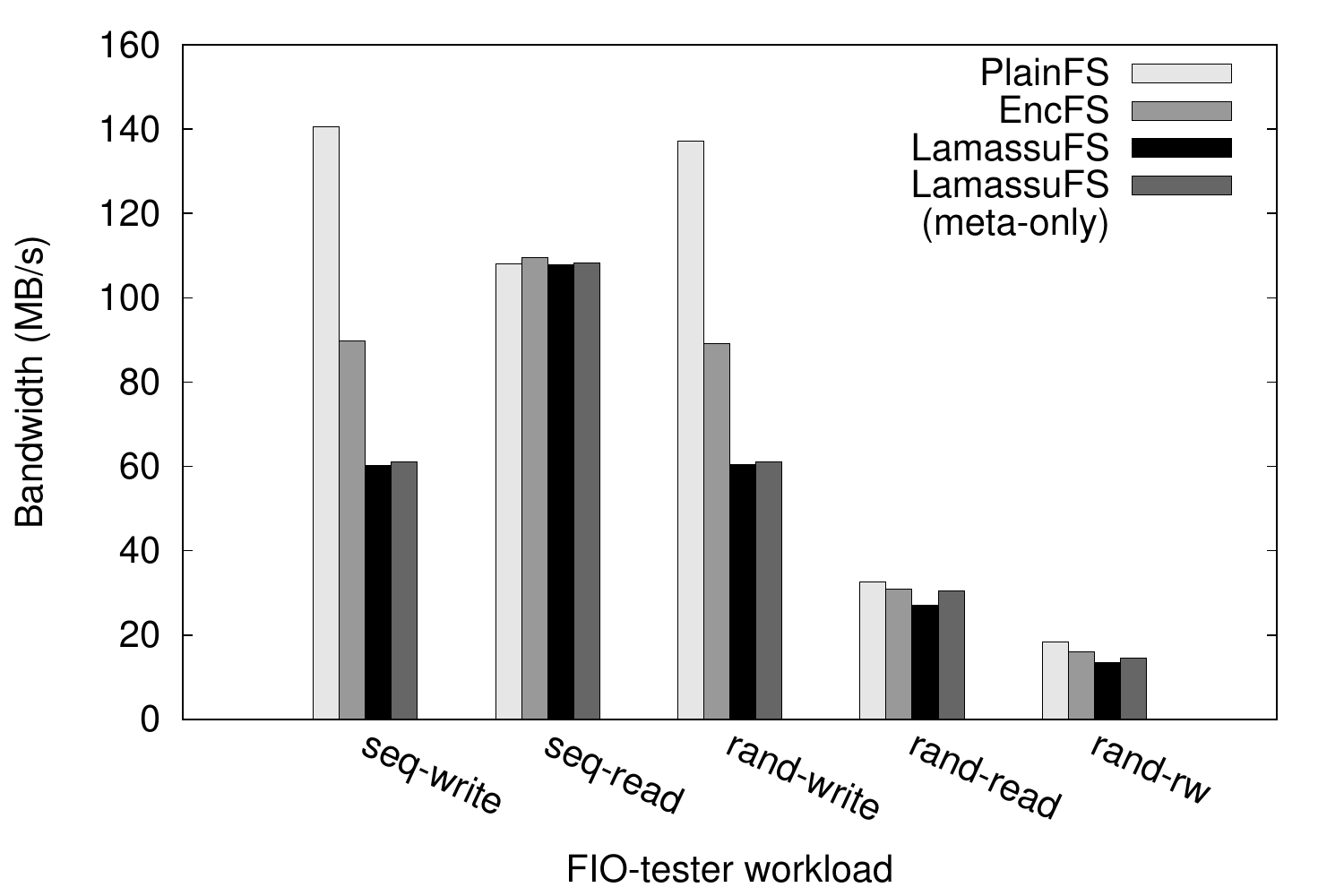, scale=.5, angle=0}%width=\textwidth}
\vspace{-.15in}
\caption{Single-file I/O throughput with a remote filer}
\label{fig:nfs_bw}
\end{figure}

Figure~\ref{fig:nfs_bw} shows the single-file I/O throughput of
PlainFS, EncFS, LamassuFS, and LamassuFS(meta-only) working with the remote filer via NFS.
With pure write workloads (seq-write and rand-write), 
we see that PlainFS performs much better than both EncFS and LamassuFS.
With pure read workloads (seq-read and rand-read),
the throughput does not show any meaningful difference across all FS:
LamassuFS shows slightly worse performance than EncFS (1.6\% to 12.4\% worse)
with read workloads.
However, LamassuFS is noticeably worse than EncFS with write workloads:
32.9\% for seq-write and 32.2\% for rand-write.

The difference in write performance is due to
per-block SHA-256 hash computations that are necessary for convergent encryption.
Because this happens at the very beginning of block encryption process,
extra latency caused by SHA-256 computation has a direct negative impact 
on I/O throughput.
On the other hand,
extra SHA-256 computation that happens during the LamassuFS read path
(for data block integrity checking)
rarely affects the performance:
LamassuFS and LamassuFS(meta-only) do not show any meaningful
throughput difference.
This suggests that 
NFS I/O is a dominant performance bottleneck in read workloads,
and therefore the rest of the computation that happens after I/O
has almost has no impact on overall I/O throughput.
This also explains why both EncFS and LamassuFS are as good as PlainFS
with read workloads.
A possible option for improving the write performance
is to increase the number of reserved key slots ($R$) in a metadata block --- with some trade-offs; 
we will discuss this later in \S\ref{subsec:varying_R}.

Overall, despite the performance overhead caused by the extra hash computation
and metadata I/O, 
we can say that the performance of LamassuFS
is competitive with that of EncFS in an NFS-shared storage environment.
Lamassu's strategy of inserting metadata blocks in a block-aligned fashion
turns out to play a significant role in terms of performance in our 
experimental environment. 
We have observed that block-unaligned accesses over NFS
incur a huge performance overhead.
For example, block-unaligned EncFS is at least 10x slower than block-aligned one when used over NFS:
7MB/s versus 85MB/s throughput in the case of seq-write. 
For this reason, to ensure as fair a comparison as possible, we have configured EncFS so that it does not add any unaligned metadata in our experiments.

\begin{figure}%[hb!]
\centering
\epsfig{file=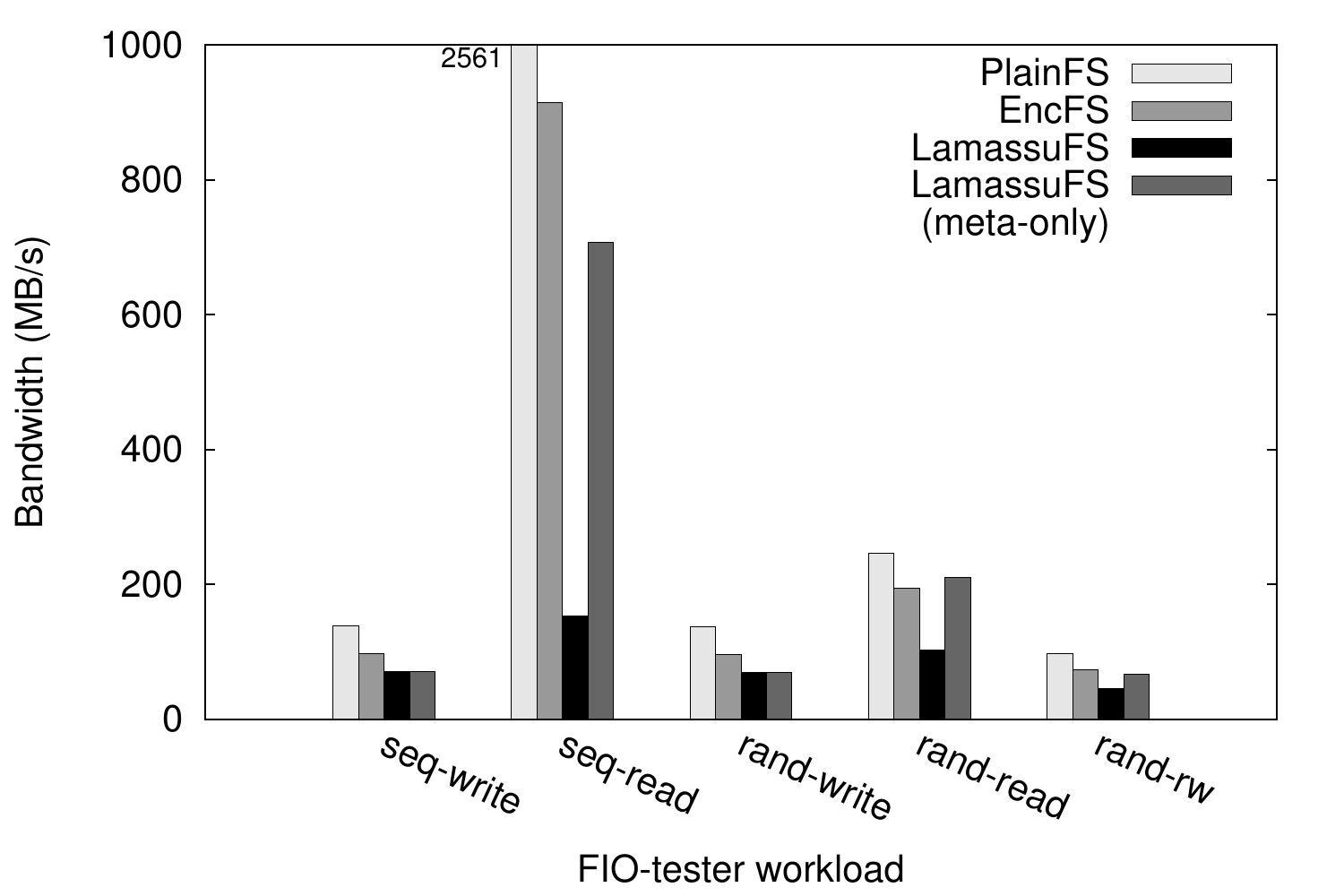, scale=.5, angle=0}%width=\textwidth}
\vspace{-.15in}
\caption{Single-file I/O throughput with a RAM disk}
\label{fig:tmpfs_bw}
\end{figure}

In order to evaluate the pure performance overhead of encryption 
without the impact of NFS I/O affected by the network bandwidth,
we ran the same set of FIO tests
by setting up all file systems so that they used
a local RAM disk (\texttt{tmpfs} in Linux) 
as backing stores, instead of the remote filer.
Figure~\ref{fig:tmpfs_bw} shows the single-file I/O throughput of all FS
with a local RAM disk.
PlainFS always noticeably performed better than EncFS and LamassuFS
across all workloads:
The difference is the greatest with seq-read
showing 2.80x over EncFS, 16.70x over LamassuFS
(note that the graph has a short y-axis).

After removing the NFS I/O bottleneck from the read path,
computation that occurs after I/O becomes a dominant bottleneck.
In particular, extra SHA-256 hash computation
that is added for a data block integrity check
negatively affects the read throughput of LamassuFS significantly:
LamassuFS performs 83.2\% worse than EncFS with a full data integrity check,
but it performs only 22.8\% worse without it.
LamassuFS also performs worse than EncFS with the rand-read workload:
47\% worse than EncFS.
However, it shows a slightly better (8.1\%) throughput than EncFS 
without full data integrity checking (meta-only): 
This is due to a small amount of write buffering introduced 
to provide consistency and reduce overall I/O.
(Recall that $R=8$ in these experiments.)
The seq-write and rand-write results are quite similar to those of NFS cases:
EncFS performs 26.8\% to 27.5\% better than LamassuFS with write workloads.

\begin{figure}%[hb!]
\centering
\epsfig{file=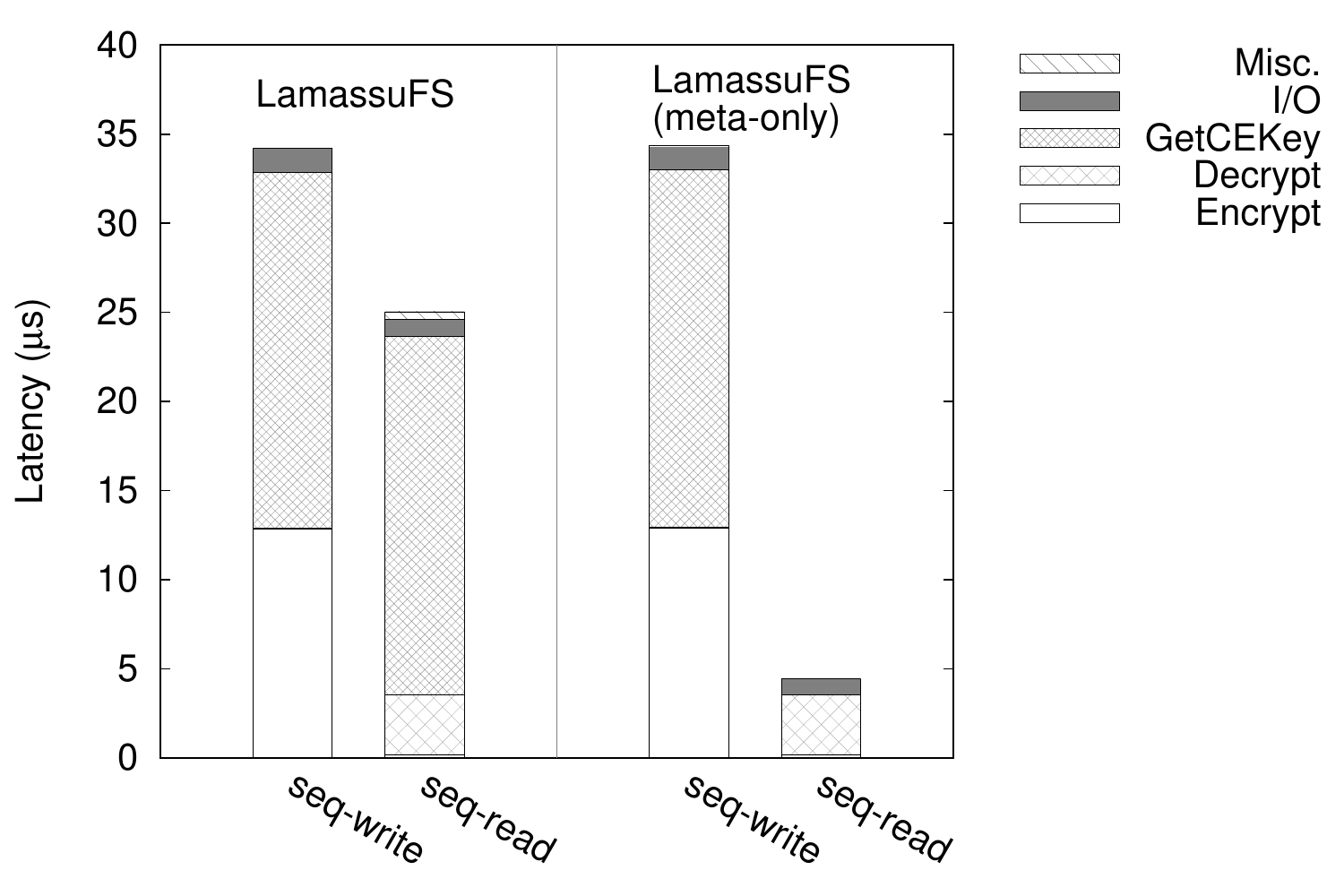, scale=.5, angle=0}%width=\textwidth}
\vspace{-.15in}
\caption{Write/read latency breakdown of LamassuFS with, and without, a full data integrity check, on a RAM disk}
\label{fig:tmpfs_bd}
\end{figure}

In order to evaluate 
the impact of SHA-256 hash computation on performance,
we broke down the write and read latency
of LamassuFS when it operates on a local RAM disk.
By inserting the instrumentation code,
the time spent on the LamassuFS read or write path
is measured and divided into five categories:
Encrypt, Decrypt, GetCEKey, I/O and Misc.
Note that the major workload of GetCEKey is SHA-256 hash computation.
Figure~\ref{fig:tmpfs_bd} shows seq-write and seq-read latency breakdown
of LamassuFS with, and without, full data integrity check.
For both writes and reads, GetCEKey consumes the most time:
58\% of seq-write, 80\% of seq-read latency.
Without full data integrity check (meta-only),
the read latency reduces drastically (81\%)
because SHA-256 hash computation is not on the read path.

There are a couple of possible options to improve the performance of LamassuFS. 
Since we have identified that the SHA-256 hash computation is the biggest performance bottleneck,
the first option is using a different cryptographic hash function
that consumes fewer CPU cycles.
For example, our microbenchmark results showed that 
OpenSSL SHA-1 consumes 58\% fewer,
and OpenSSL MD5 consumes 38\% fewer CPU cycles 
for computing the same 4KB block-hash
compared with our SHA-256 function using the Intel AVX instruction set.
The exact implication of using a less secure hash function
(e.g., SHA-1 or MD5 generates 128-bit keys instead of 256-bit keys) 
for convergent encryption
could be understood only with comprehensive cryptographic analysis,
and hence we will leave it for future work.

The second option is to forgo data block integrity check in the read path. 
This will improve the read performance significantly, as shown in
Figure~\ref{fig:tmpfs_bw} with LamassuFS(meta-only).
Remember that Lamassu is still doing metadata block integrity checking via AES-GCM:
It is always able to detect any data corruption
that occurs within or across metadata blocks
(e.g., first 4KB block from one hundred and nineteen 4KB blocks if $R=8$).
This covers cases such as accidentally overwriting a whole file, or
the beginning of a file.
In an enterprise storage environment with no malicious user
who intentionally corrupts the user data,
this could be a viable option for improving performance 
while sacrificing a little on security.

\subsection{Number of Reserved Key Slots}
\label{subsec:varying_R}

As described in \S\ref{subsec:consistency},
Lamassu maintains a certain number of reserved key slots (denoted $R$)
in a metadata block to maintain consistency.
As $R$ increases, the amount of space taken up by Lamassu metadata in each file increases, and thus storage efficiency decreases.
On the other hand,
increasing $R$ reduces the number of additional metadata I/Os that Lamassu must perform. 
To maintain consistency, Lamassu caches block writes in memory and writes 
them to disk along with their metadata as part of a commit operation. 
That occurs once for every $R$ data block writes.
The decreased number of writes positively affects the write throughput,
while the increased metadata space overhead negatively affects storage efficiency. 
In order to understand this trade-off,
we evaluated the performance and storage efficiency of LamassuFS
by varying $R = $  
1, 2, 8, 32, 48, 52, 56, and 60.

\begin{figure}%[ht]
\centering
\epsfig{file=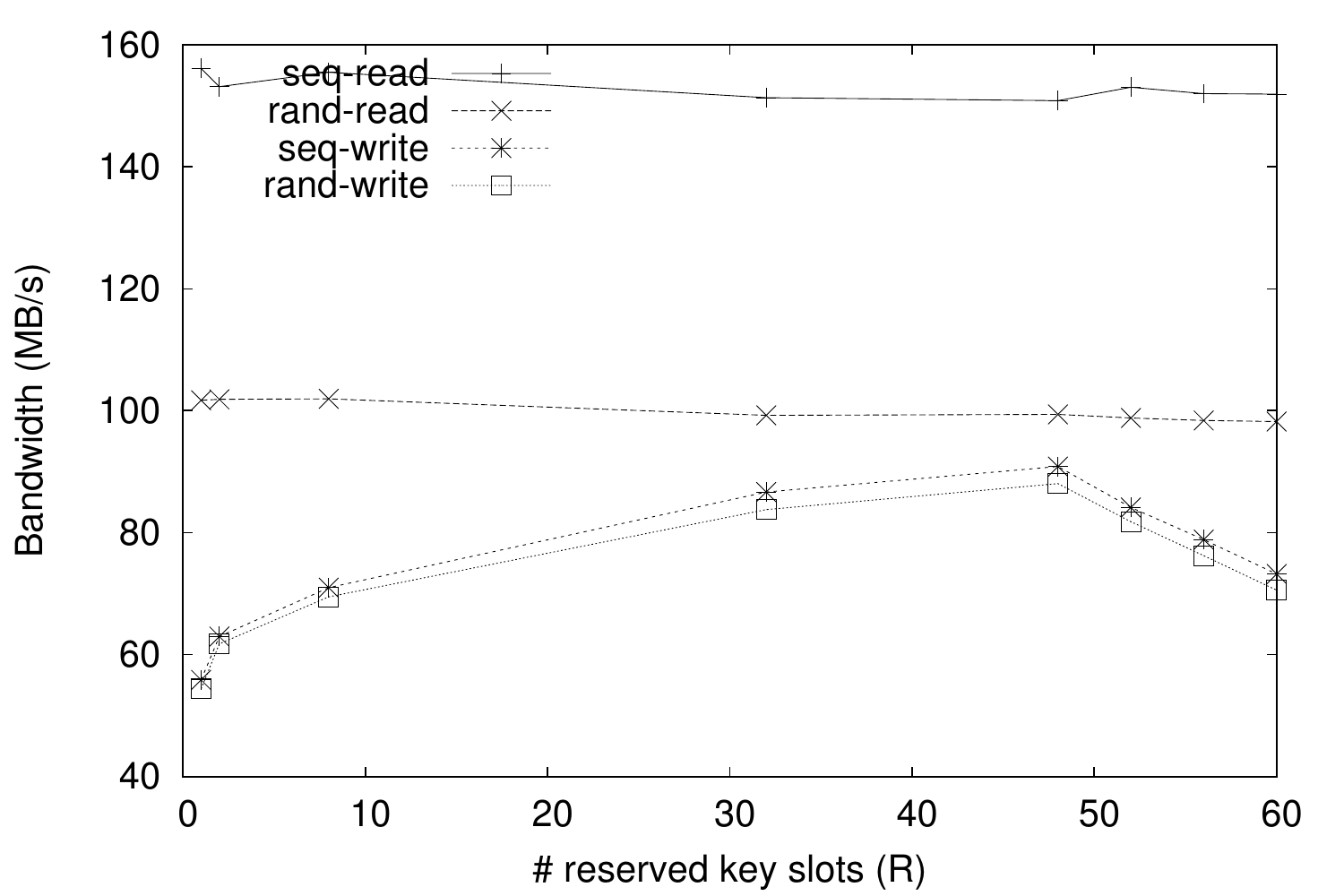, scale=.48, angle=0}%width=\textwidth}
\vspace{-.2in}
\caption{Single-file I/O throughput by varying $R$}
\label{fig:r_bw_tmpfs}
\vspace{.1in}
\centering
\epsfig{file=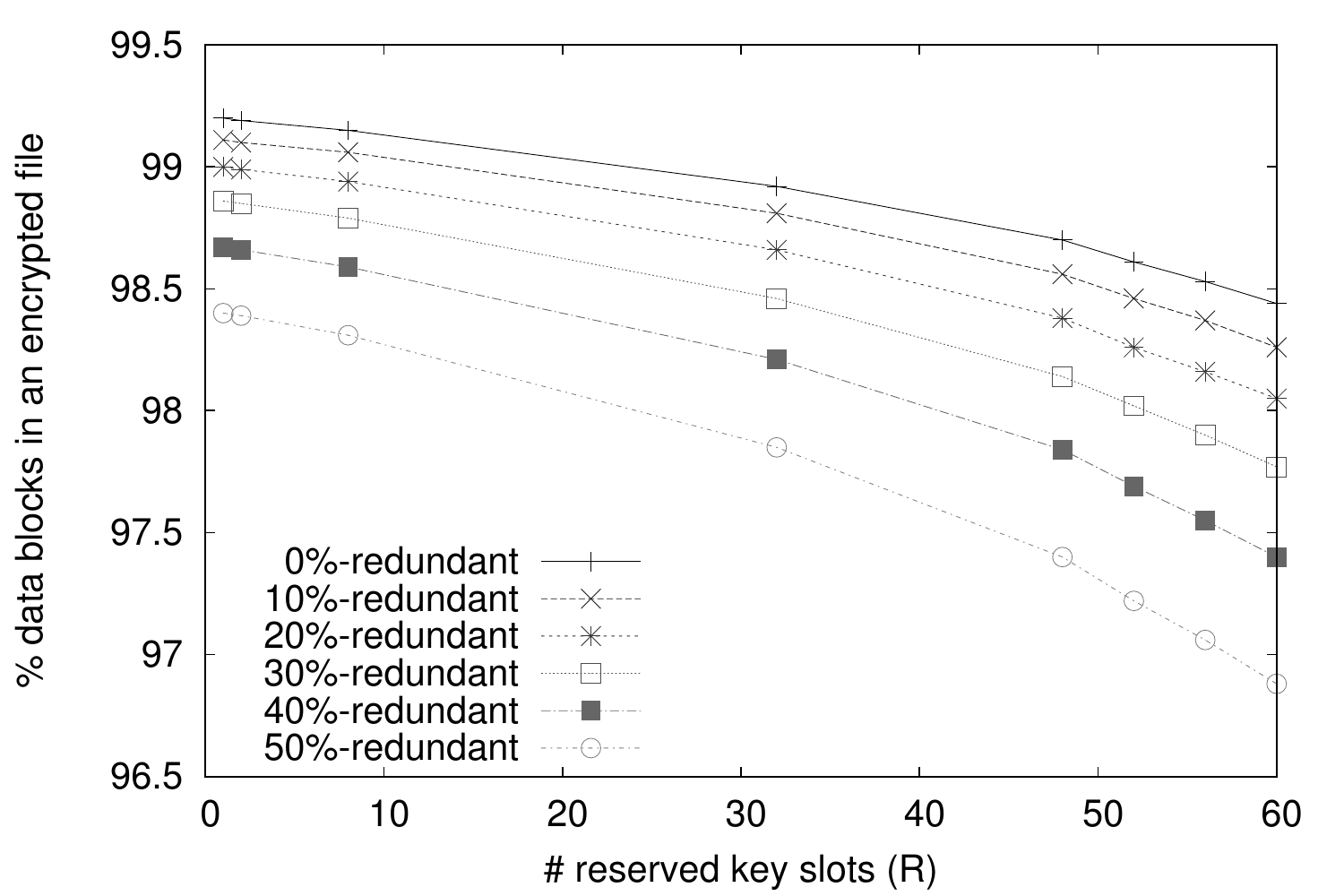, scale=.48, angle=0}%width=\textwidth}
\vspace{-.2in}
\caption{Storage efficiency by varying $R$}
\label{fig:r_se}
\end{figure}

Figure~\ref{fig:r_bw_tmpfs} shows the single-file I/O throughput in bandwidth
when applying four different FIO-tester workloads to LamassuFS with a local RAM disk back-end.
By increasing $R$, 
the write throughput continuously improves up to a certain point and then decreases:
The throughput reaches its peak around $R=48$ achieving 
1.60x and 1.57x speedups over $R=1$ for seq-write and rand-write 
respectively. 
For write workloads,
the positive impact of buffering and batching of writes 
(i.e., reduced metadata I/O)
is a dominant factor
as it increases the write throughput significantly.
On the other hand,
the read throughput tends to decrease slightly as $R$ increases:
from 1 to 60, 4.71\% and 4.40\% decreases for seq-read and rand-read, respectively.
This is
because Lamassu must read more metadata blocks per
unit file size with a larger $R$ value 
(i.e., metadata space overhead increases), 
resulting in a slight increase in I/O overhead.

Figure~\ref{fig:r_se} shows storage efficiency as
the percentage of data blocks, excluding metadata blocks, 
in different encrypted files with various redundancy profiles 
(denoted as $\alpha$ in \S\ref{subsec:res_se}).
The storage efficiency decreases as $R$ increases 
because of the larger space overhead of metadata.
The storage efficiency also decreases as there are more redundant blocks in a plaintext file (i.e., $\alpha$ increases)
because the metadata blocks are not deduplicated,
as previously shown in \S\ref{subsec:res_se}.

The right value of $R$ should be chosen with consideration for this trade-off.
If an application requires higher write IOPS,
a larger $R$ can be chosen
while sacrificing a little space efficiency.
However, with larger $R$, the granularity of crash consistency becomes coarser,
(i.e., it increases the recovery point objective [RPO] of the system.),
and LamassuFS consumes the additional memory space for more write buffers.
For the proceeding set of experiments, 
we fixed $R$ to be 8 to achieve a balanced trade-off.

\section{Related Work}
\label{sec:relwork}

\subsection{Encrypted File Systems}

Full disk encryption (FDE)
is a popular choice for the storage encryption.
NetApp Storage Encryption (NSE)~\cite{NSE:Online}
is a hardware-based implementation of FDE
that uses self-encrypting drives (SEDs) from drive vendors. 
FileVault 2~\cite{FileVault2:Online} is a
software-based FDE in Mac OS X
that uses Intel's AES-NI.
As encryption occurs at the lowest stack
just before data blocks are written to the disk,
FDE is quite a different approach from
our data-source encryption strategy.

FDE encrypts whole blocks in a volume,
including file system metadata, 
while file-system-level encryption enables
encryption of individual files or directories,
offering a finer granularity of control.
There are general-purpose file systems that
have integrated encryption features, 
such as ZFS and Encrypting File System (EFS) in NTFS.
Some cryptographic file systems
--- such as
CFS~\cite{Blaze:1993:CFS:168588.168590},
TCFS~\cite{Cattaneo:2001:DIT:647054.715628},
Cryptfs~\cite{Zadok98cryptfs:a}
and eCryptfs~\cite{Halcrow:eCryptfs} ---
are stackable on top of another general-purpose file system;
eCryptFS is widely used, included in Ubuntu's encrypted home directory
and Google's Chrome OS.
There are a few FUSE-based encrypted file systems available;
only EncFS~\cite{EncFS:Online} is notable in terms of its maturity and wide acceptance.

With regard to our data-source encryption strategy,
any stackable or FUSE-based encrypted file systems
can serve the same purpose
because they can transparently work on top of an existing system.
However, to the best of our knowledge,
none of them provides the explicit ability to 
enable deduplication at downstream storage devices
as Lamassu does.
We chose a FUSE-based file system for our prototype implementation
due to its better debuggability and easier deployment;
it can also be implemented as a kernel-level file system 
if necessary.

\subsection{Convergent Encryption}

Deduplication of previously encrypted data is normally impossible
because of the nature of encrypted data. Convergent encryption (CE) 
was proposed to address this issue.
The concept of convergent encryption was introduced by 
Douceur et al.~\cite{DouceurABST02}:
By definition, CE produces identical ciphertext files from identical plaintext files. 
A common approach is deriving the encryption key from a secure hash of the plaintext.
Using convergent encryption results in the ciphertext having the same levels of duplication as the plaintext. 
However, the weakness of CE is the leakage of information about the plaintext; 
an attacker can observe the ciphertext 
and deduce the contents of the plaintext by using a variety of different attacks~\cite{Zooko:2009}.

While the system described by Douceur, et al.\cite{DouceurABST02} works only with the whole files,
Storer et al.~\cite{Storer:2008:SDD} later designed a CE solution
that provides sub-file granularity encryption and deduplication, in both fixed and variable sized chunks. 
Bellare et al.~\cite{BellareKR12} formalized CE
as Message-Locked Encryption (MLE)
with a cryptographic analysis.
DupLESS~\cite{Keelveedhi:2013} tried to overcome CE's weakness
--- leakage of information about the plaintext ---
with an obfuscated key exchange mechanism with a key server
in order to achieve stronger confidentiality;
however, the performance overhead turns out to be quite costly
as it requires 3-way key exchange with a key server 
for every block access.
ClouDedup~\cite{Puzio:2013} uses
a semi-trusted server between users and the cloud provider
to encrypt the ciphertext resulting from CE with another encryption algorithm,
and a metadata manager to store encryption keys and block signatures.
It introduces much more complexity to the system, compared to Lamassu, 
and might incur performance penalty due to double encryption.

Lamassu is targeted at enterprise environments
in which multiple hosts store data in a large shared storage appliance.
Therefore, it tries to achieve a balance between performance and security.
Lamassu can be easily added to an existing enterprise environment. It only incurs 
a mimal performance overhead, and does not require an extra system for a metadata store.
In a multitenant system, tenant data can be securely separated by using per-tenant
keys to create isolation zones.
Tahoe-LAFS~\cite{Wilcox-O'Hearn:2008:TLF:1456469.1456474}
used a similar approach of adding a secret during hash key generation~\cite{Zooko:2009},
but its convergent encryption works on a per-file basis, 
limiting the storage efficiency
compared with Lamassu's per-block approach.

\section{Conclusion}
\label{sec:conc}

In this paper, we presented Lamassu, a new, transparent, encryption system
that provides strong data-source encryption,
while preserving downstream storage-based data deduplication.
Lamassu uses block-oriented convergent encryption 
to align with existing block-based deduplication systems. 
It takes a new approach to manage convergent encryption key metadata 
by inserting it into each file's data stream, eliminating the need for
additional infrastructure. 
Therefore, it can be inserted into an existing application stack 
without any modification to either host-side applications
or the storage controller.
We also introduced a strategy for maintaining consistency between 
file data and convergent metadata, and 
for providing data integrity checking for application data in a 
convergent encryption system. 

Our results showed that it is possible to insert convergent encryption 
into an application with a performance overhead 
similar to non-convergent options, 
placing encryption near the top of the application stack, 
and making it easier to provide strong security across the whole stack. 
Our security model leaks only the information 
that is absolutely necessary for deduplication to the storage system, 
resulting in strong encryption, well suited to many applications. 
Our system provides a clear advantage over analogous solutions 
by preserving storage-based deduplication without compromising on encryption.

\paragraph{Acknowledgments.}
We thank Emil Sit, ATC PC members and anonymous reviewers for their feedback and guidance.
We also thank James Kelley and James Lentini at NetApp for their help and support for this work.

{\footnotesize \bibliographystyle{acm}
\bibliography{dedup}}

\footnotesize
\noindent
NetApp, the NetApp logo and Data ONTAP are trademarks, or registered trademarks, of NetApp, Inc. in the United States and/or other countries. All other brands or products are trademarks, or registered trademarks, of their respective holders, and should be treated as such. A current list of NetApp trademarks is available on the web at \url{http://www.netapp.com/us/legal/netapptmlist.aspx}.

\end{document}